\documentclass[12pt]{iopart}
\usepackage{iopams}  
\usepackage{url}

\newcommand{\cL}{{\mathcal L}}
\newcommand{\cD}{{\mathcal D}}
\newcommand{\cM}{{\mathcal M}}
\newcommand{\cJ}{{\mathcal J}}
\newcommand{\cC}{{\mathcal C}}
\newcommand{\mbar}{\bar{m}}

\begin{document}

\title[Optomechanical damping basis]{Optomechanical damping basis}

\author{Juan Mauricio Torres}
\address{Instituto de F\'isica, Benem\'erita Universidad Aut\'onoma de Puebla, Apdo. Postal J-48, Puebla 72570, M\'exico}

\author{Ralf Betzholz}
\address{School of Physics and International Joint Laboratory on Quantum Sensing and Quantum Metrology, Huazhong University of Science and Technology, Wuhan 430074, China}	
\ead{ralf\_betzholz@hust.edu.cn}

\author{Marc Bienert}
\address{Sch\"ulerforschungszentrum S\"udw\"urttemberg, D-88348 Bad Saulgau, Germany}

\vspace{10pt}
\begin{indented}
\item[]\today
\end{indented}

\begin{abstract}
We present a closed-form analytical solution to the eigenvalue problem of the Liouville operator 
generating the dissipative dynamics of the standard optomechanical system. The corresponding Lindblad master
equation describes the dynamics of a single-mode field inside an optical cavity coupled
by radiation pressure to its moving mirror. The optical field and the mirror are in contact with separate environments, which are assumed at zero and finite temperature, respectively. The optomechanical damping basis refers to the exact set of eigenvectors of the generator 
that, together with the exact eigenvalues, are explicitly derived. Both the weak- and the strong-coupling
regime, which includes combined decay mechanisms, are solved in this work.
\end{abstract}

\section{Introduction}
Exact analytical solutions are rare in quantum mechanics.
Even for closed systems the number of solvable problems is rather limited, and yet fewer solutions are known when this condition is relaxed. 
For open systems in contact with a Markovian environment the quantum dynamics is well described by Lindblad master equations~\cite{Lindblad1976,Gorini1976}. 
Solving this type of equations is not an easy task and typically one has to rely on numerical and
perturbative methods, while for many applications one may be satisfied with finding the stationary state or some of its properties~\cite{Prosen2011,Karevski2013,Asadian2013,Prosen2014,Albert2014,Bernad2015,Guo2017,Ribeiro2018}. 
A full solution can be obtained by
the spectral decomposition of the non-Hermitian Liouville superoperator generating the 
dynamics~\cite{Briegel1993,Barnett2000,Prosen2008,Prosen2010}.  
Once the eigenvalue problem is solved, one can, in principle, evaluate the time-dependent 
density operator of the system for any given initial condition~\cite{Englert2002,Jakob2003,Maniscalco2006,Betzholz2014,Torres2017}, 
or the eigensystem can serve as the basis for a perturbative treatment of more involved problems~\cite{Barnett2001,Bienert2004,Bienert2007,Betzholz2016,Li2014}.
Among the few solvable instances found in the literature, 
a prominent example is the damped harmonic oscillator ~\cite{Briegel1993,Barnett2000,Barnett2001,Englert2002}. 
It was in the context of its solvability, that the concept of the damping basis as 
the set of eigenvectors of the dissipative generator was introduced in Ref.~\cite{Briegel1993} for the 
treatment of an atom-field system in quantum optics. However, applications and extensions have not gone much further since~\cite{Torres2014,Betzholz2014}.  

Beyond the damped harmonic oscillator, the standard optomechanical system
stands out as an interesting nonlinear open quantum model that has drawn considerable attention of the community \cite{Aspemayer2014,Kippenberg2008}. 
Inspired by astonishing achievements in the control of single atomic systems through the light-matter interaction in quantum optics~\cite{Leibfried03,Meschede2006},
the possibility to manipulate macroscopic objects
using radiation pressure has been explored~\cite{Ashkin}. Quantum control of truly macroscopic
objects is not only interesting for applications, such as displacement and force detection near the fundamental quantum limit~\cite{Caves1980,Braginsky,Gavartin2012}, but also from the
point of view of testing the foundations of quantum mechanics on a larger scale than
the atomic one~\cite{Chen2013}. Here, the paradigm is a cavity optomechanical setup, consisting of a Fabry-P\'erot cavity with one fixed and one harmonically suspended mirror,
in which the light field of the optical resonator exerts a radiation-pressure force on the moving mirror~\cite{Meystre2013,Aspemayer2014,Bowen,Metcalfe2014}.
Tremendous progress toward full quantum control over the mechanical degree of
freedom in such systems has been achieved, for example by the demonstration of laser cooling of
its motion toward the ground state~\cite{Chan2011,Teufel2011}. 
A theoretical approach for treating the dissipative nature of the optomechanical system is a 
Lindblad master equation describing the dynamics of two coupled harmonic 
oscillators~\cite{Marquardt2008,Wilson-Rae2008b}. 
Remarkably, the standard optomechanical system 
without coherent driving of the cavity, 
including the strong-coupling limit, is an exactly solvable open quantum system.

In this paper, we solve the eigenvalue problem of the Liouville superoperator generating the
dynamics of a typical dissipative optomechanical system in an exact form.
We show that the eigenvalues are given as the
sum of the eigenvalues of the uncoupled systems plus a term that arises due to their interaction that
can be written in closed form.  Extending the definition in Ref.~\cite{Briegel1993}, we introduce
the optomechanical damping basis as the exact right and left eigenvectors of the non-Hermitian generator, which are explicitly derived.

\section{Optomechanical master equation}
The unitary dynamics of the standard optomechanical system is governed by the Hamiltonian
\begin{eqnarray}
H=\hbar\omega a^\dagger a+\hbar\nu b^\dagger b-\hbar\chi a^\dagger a(b+b^\dagger),
\label{Hamiltonian}
\end{eqnarray}
where the first part describes the energy of the single-mode optical cavity with frequency $\omega$ in terms of the annihilation operator $a$ and the second one accounts for the
harmonic motion of the mirror with frequency $\nu$ and annihilation operator $b$. The last term represents the optomechanical interaction~\cite{Law1994,Law1995} with strength $\chi$.

For weak optomechanical coupling the dissipative dynamics of the density operator $\rho$ obeys the Lindblad master equation $\partial\rho/\partial t=\cL\rho$ with the Liouville superoperator
\begin{eqnarray}
\cL=\cL_{\rm C}+\cL_{\rm M}+i\chi\cC[a^\dagger a(b+b^\dagger)].
\label{master}
\end{eqnarray}
The first term describes the lossy optical cavity in contact with a zero-temperature bath and reads
\begin{eqnarray}
\cL_{\rm C}=-i\omega \cC[a^\dagger a]+\frac{\kappa}{2}\cD[a],
\end{eqnarray}
where $\kappa$ is the linewidth and we have written the coherent part and the cavity decay in terms of a commutator and a Lindbladian defined as
\begin{eqnarray}
\cC[X]\rho=[X,\rho],\\
\cD[X]\rho=2X\rho X^\dagger-X^\dagger X\rho-\rho X^\dagger X.
\end{eqnarray}
The zero-temperature assumption is justified in this case since the mean thermal photon number is negligible for optical frequencies.
The second contribution in~\eref{master} accounts for the dynamics of the mechanical oscillator, including damping by a heat bath at finite temperature $T$, and reads
\begin{eqnarray}
\cL_{\rm M}=-i\nu\cC[b^\dagger b]+\frac{\gamma(\mbar +1)}{2}\cD[b]+\frac{\gamma \mbar}{2}\cD[b^\dagger],
\end{eqnarray}
where $\gamma$ is the  damping rate and the mean thermal phonon number is given by $\mbar=[\exp(\hbar\nu/k_{\rm B}T)-1]^{-1}$. Here, we briefly mention that by writing the Liouvillian in the form~\eref{master} we have added the cavity decay and the damping of the harmonic oscillator independently. This assumption is valid in the case of a relatively weak optomechanical coupling $\chi\ll \nu$, which is given in most experimental realizations~\cite{Aspemayer2014}. The ultra-strong coupling regime, in which this condition is not fulfilled anymore, is treated in section~\ref{strong}. In that case, while the two decay mechanisms can no longer be treated independently, the form of optomechanical interaction in~\eref{Hamiltonian} still remains applicable.

The Liouvillian in~\eref{master} can be rewritten as
\begin{eqnarray}
\cL=\cM+\cJ,\\
\cJ\rho=\kappa a\rho a^\dagger,
\end{eqnarray}
where we introduced  the photonic jump superoperator $\cJ$.
In this way we have separated $\cL$ in two parts. One that only contains the number
operator $a^\dagger a$ and conserves the photon number, namely the superoperator $\cM$. The remaining part $\cJ$ connects adjacent subspaces with definite photon numbers. This separation will prove useful in solving the eigensystem of this model~\cite{Torres2014}.

The formal solution to the master equation~\eref{master} can be obtained 
by diagonalization of the Liouvillian~\cite{Briegel1993}, i.e., by solving the eigenvalue problem
\begin{eqnarray}
\label{eigenvalueequation}
\cL\hat{\rho}_\lambda=\lambda\hat{\rho}_\lambda,\\
\cL^\dagger\check{\rho}_\lambda=\lambda^\ast\check{\rho}_\lambda.
\end{eqnarray}
In this notation the hats ($\hat\rho$) and checks ($\check\rho$) on the operators respectively denote the eigenvectors of $\cL$ and its adjoint $\cL^\dagger$, which we will also refer to as right and left eigenvectors. This dual system of eigenvalue equations arises due to the non-Hermiticity of $\cL$,
which implies that the diagonalization transformation is not unitary. 
For a superoperator $\cL$ with a non-degenerate spectrum the left and right eigenvectors can be normalized to fulfill the orthonormality condition with respect to the Hilbert-Schmidt inner product, viz.
\begin{eqnarray}
\label{orthogonality}
{\rm Tr}\big\{\check{\rho}_\lambda^\dagger\hat{\rho}_{\lambda'}\big\}=\delta_{\lambda,\lambda'}.
\end{eqnarray}
The explicit form of $\cL^\dagger$ can be derived 
by noting that the commutator $\cC$ is self-adjoint and
from the 
inner product according to $\Tr\{X^\dagger(\cL Y)\}=
\Tr\{(\cL^\dagger X)^\dagger Y\}$.
In this way, one finds that the adjoint dissipator reads
\begin{eqnarray}
\label{adjointD}
\cD^\dagger[X]\rho=2X^\dagger \rho X-X^\dagger X\rho-\rho X^\dagger X.
\end{eqnarray}
In order to explicitly solve the eigenvalue problem of $\cL$, we will first focus on the eigensystem of $\cM$. Then we will investigate the action of the jump 
operator $\cJ$ on the eigenvectors of $\cM$. With this knowledge, a plausible ansatz
for the eigenvectors of the complete superoperator $\cL$ leads to a first order
recursion that can be solved.

\section{Diagonalization of $\cM$}
\label{diagM}
We first consider the right eigenvectors. Taking the properties of $\cM$ into account, we consider operators of the type
$|n+l\rangle\langle n\vert\mu$,
where $|n\rangle$ represents a state of $n$ photons and $\mu$ is an arbitrary mechanical-oscillator operator. The action of $\cM$ onto this class of states is
\begin{eqnarray}
\cM|n+l\rangle\langle n|\mu=|n+l\rangle\langle n\vert\cM^{(l,n)}\mu,\\
\label{Mnl}
\cM^{(l,n)}=
\cL_{\rm M}+in\chi\cC[b+b^\dagger]+il\chi(b+b^\dagger)+\lambda_{\rm C}^{(l,n)},
\end{eqnarray}
with the eigenvalues of the Liouvillian $\cL_{\rm C}$ given by 
\begin{eqnarray}
\lambda^{(l,n)}_{\rm C}=-il\omega-(n+|l|/2)\kappa,
\label{lambdaC}
\end{eqnarray}
for nonnegative integers $n$ and $l$.
Here, the last two terms in $\cM^{(l,n)}$ represent common multiplications with an operator and a complex number. For $l=0$, this corresponds to the master equation of a damped harmonic oscillator with a driving of strength $n\chi$,
which can be readily solved by displacing the solution of the undriven version by $n\chi/(\nu-i\gamma/2)$. A similar approach has been considered in Refs.~\cite{Restrepo2014,Ventura}.
Although for nonzero values of
$l$ the task seems more intricate, it is still possible to find its solution by 
considering an asymmetric displacement
\begin{eqnarray}
\tilde\mu=e^{\eta_lb}D(\alpha_{l,n})\mu D^\dagger(\beta_{l,n})e^{-\eta_lb},
\label{asdis}
\end{eqnarray}
where the displacement operator of the mechanical oscillator is given by
$D(\alpha)=\exp(\alpha b^\dagger-\alpha^\ast b)$
and with the shorthands
\begin{eqnarray}
\beta=\frac{\chi}{\nu-i\gamma/2},\quad
\alpha_{l,n}=-(n+l)\beta-il|\beta|^2\gamma\mbar/\chi,\\
\beta_{l,n}=\alpha_{l,n}+l\beta^\ast,\quad
 \eta_l=il|\beta|^2\gamma(2\mbar+1)/\chi.
\label{alfabetas}
\end{eqnarray}
Applying the transform~\eref{asdis} to~\eref{Mnl} one finds
\begin{eqnarray}
\label{Mtilde}
\tilde\cM^{(l,n)}\tilde\mu=\left[\cL_{\rm M}+\lambda_{\rm C}^{(l,n)}+\varepsilon_{l,n}\right]\tilde\mu,\\
\varepsilon_{l,n}=l|\beta|^2\left[i(2n+|l|)\nu-l\gamma(\mbar+1/2)\right].
\label{epsilon}
\end{eqnarray}
This is nothing but the Liouvillian $\cL_{\rm M}$ of a damped harmonic oscillator, apart from two constant terms, whose eigensolutions are known~\cite{Briegel1993,Barnett2000,Englert2002}.
The corresponding eigenvalues of $\cL_{\rm M}$ are given by
\begin{eqnarray}
\lambda^{(k,m)}_{\rm M}=-ik\nu-(m+|k|/2)\gamma,
\end{eqnarray}
with integer $k$ and nonnegative integer $m$ ($k\in \mathbb{Z}$, $m\in \mathbb{N}$).
As for the eigenvectors, they can be written in the form
\begin{eqnarray}
\label{mu_right}
\hat{\mu}_{k,m}=\frac{1}{(\mbar+1)^{k+1}}b^{\dagger k}\left\{L_m^{(k)}\left(\frac{b^\dagger b}{\mbar+1}\right)e^{-\frac{b^\dagger b}{\mbar+1}}\right\}_{\rm n},\\
\label{mu_left}
\check{\mu}_{k,m}=\frac{m!}{(m+k)!}\left\{L_m^{(k)}\left(\frac{b^\dagger b}{\mbar+1}\right)\right\}_{\rm a}b^{\dagger k},
\end{eqnarray}
for $k\geq0$, while for $k<0$ one has to take the Hermitian conjugate expressions, with $k$ replaced by $|k|$. Here, $\{\cdot\}_{\rm n}$ and $\{\cdot\}_{\rm a}$ respectively stand for normal and antinormal ordering~\cite{Cahill1969} and $L_m^{(k)}$ denotes Laguerre polynomials~\cite{Gradshtyn}. The
eigenvectors of $\cM^{(l,n)}$ are thereby
\begin{eqnarray}
\label{righteigenM}
\hat{\mu}_{k,m}^{(l,n)}=D^\dagger(\alpha_{l,n})e^{-\eta_lb}\hat{\mu}_{k,m}e^{\eta_lb}D(\beta_{l,n})
\end{eqnarray}
and in terms of these mechanical operators the eigenvalue problem for $\cM$ is solved by
\begin{eqnarray}
\cM|n+l\rangle\langle n|\hat{\mu}_{k,m}^{(l,n)}=\lambda_{k,m}^{(l,n)}|n+l\rangle\langle n|\hat{\mu}_{k,m}^{(l,n)},
\label{eigenvalueeqM}\\
\lambda_{k,m}^{(l,n)}=\lambda_{\rm M}^{(k,m)}+\lambda_{\rm C}^{(l,n)}+\varepsilon_{l,n}.
\label{eigenvaluesM}
\end{eqnarray}
We mention that a degeneracy of the eigenvalues can only occur for rare choices of parameters, which we do not assume in the following. 
The eigenvectors of $\cM$ containing the cavity operators $|n\rangle\langle n+l|$ can be obtained by Hermitian conjugation of~\eref{eigenvalueeqM}.

Let us now turn to the left eigenvectors. Using~\eref{adjointD} one finds that the adjoint of $\cM^{(l,n)}$ is given by 
\begin{eqnarray}
\cM^{\dagger(l,n)}=
\cL^\dagger_{\rm M}
+in\chi\cC[b+b^\dagger]-il\chi(b+b^\dagger)+\lambda_{\rm C}^{\ast(l,n)}.
\end{eqnarray}
The solution of the eigenvalue equation of $\cM^{\dagger(l,n)}$ is also achieved by an asymmetric displacement, analogous to~\eref{asdis}, and the resulting left eigenvectors are given by $|n+l\rangle \langle n|\check{\mu}_{k,m}^{(l,n)}$ with the mechanical operators
\begin{eqnarray}
\check{\mu}_{k,m}^{(l,n)}=D^\dagger(\alpha_{l,n})e^{-\eta_lb^\dagger}\check{\mu}_{k,m}e^{\eta_lb^\dagger}D(\beta_{l,n}).
\end{eqnarray}

\section{Full diagonalization}
\label{diagFull}
We now use the eigensystem of $\cM$ to establish the eigensystem of the full Liouvillian by including the jump superoperator. Noting that $\cJ$ couples only elements on the same diagonal in the photonic basis, 
we choose the following ansatz
\begin{eqnarray}
\label{ansatz}
\hat{\rho}_{k,m}^{(l,n)}=\sum_{j=0}^{n}|j+l\rangle\langle j\vert\hat{\varrho}_{k,m}^{(l,n;j)}
\end{eqnarray}
for the right eigenvectors, where the mechanical operators in this expansion have to be determined. Inserting this ansatz in ~\eref{master}, one obtains for the different 
subspaces $|j+l\rangle\langle j\vert$ the following equalities
\begin{eqnarray}
\label{expansion1}
\left(\lambda_{k,m}^{(l,n)}-\cM^{(l,n)}\right)\hat\varrho_{k,m}^{(l,n;n)}=0,
\\
\label{expansion2}
\left(\lambda_{k,m}^{(l,n)}-\cM^{(l,j-1)}\right)
\hat\varrho_{k,m}^{(l,n;j-1)}
=\kappa\sqrt{j(j+l)}\hat\varrho_{k,m}^{(l,n;j)}.
\end{eqnarray}
Equation~\eref{expansion1} has two important implications. One is  that the mechanical part of the $n$th term 
of the expansion~\eref{ansatz} is an eigenvector
of the superoperator $\cM^{(l,n)}$, namely $\hat\varrho_{k,m}^{(l,n;n)}=\hat\mu_{k,m}^{(l,n)}$.
The second is that the eigenvalues of the full
Liouvillian are the same as those of $\cM$, namely $\lambda=\lambda_{k,m}^{(l,n)}$, which means that $\cL$ is already triangular in the eigenbasis of $\cM$. 
Equation~\eref{expansion2} on the other hand shows that the operators in the expansion~\eref{ansatz} have to fulfill a first-order recursion in order to solve the eigenvalue equation. 
With the boundary condition~\eref{expansion1} 
this recurrence is solved by
\begin{eqnarray}
\label{righteigenL}
\hat\varrho_{k,m}^{(l,n;j)}=a_{l,n,j}
\Bigg[\prod_{j'=j}^{n-1}\frac{\kappa}{\lambda_{k,m}^{(l,n)}-\cM^{(l,j')}}\Bigg]\hat{\mu}_{k,m}^{(l,n)},
\end{eqnarray}
with $a_{l,n,j}=\sqrt{n!(n+l)!/j!(j+l)!}$ and the convention $\prod_{j=1}^{N}\mathcal{X}_j=\mathcal{X}_1\mathcal{X}_2\cdots\mathcal{X}_N$. 

The left eigenvectors can be derived in a similar way. The action of the adjoint jump superoperator is given by $\cJ^\dagger\rho=\kappa a^\dagger\rho a$, i.e., it induces jumps upward in the photon number. Opposed to the finite superposition in the right eigenvectors~\eref{ansatz} for the left ones we make the ansatz
\begin{eqnarray}
\check{\rho}_{k,m}^{(l,n)}=\sum_{j=n}^{\infty}|j+l\rangle\langle j\vert\check{\varrho}_{k,m}^{(l,n;j)},
\end{eqnarray}
in which the sum extends to infinity since there is no upper bound for the photon number. In analogy to~\eref{expansion1} and~\eref{expansion2}, upon substitution into the eigenvalue equation~\eref{eigenvalueequation} of $\cL^\dagger$, this leads to a recurrence with the initial condition $\check{\varrho}_{k,m}^{(l,n;n)}=\check{\mu}_{k,m}^{(l,n)}$, which is solved by
\begin{eqnarray}
\label{lefteigenL}
\check{\varrho}_{k,m}^{(l,n;j)}=a_{l,j,n}\Bigg[\prod_{j'=0}^{j-n-1}\frac{\kappa}{\lambda_{k,m}^{\ast(l,n)}-\cM^{\dagger(l,j-j')}}\Bigg]\check{\mu}_{k,m}^{(l,n)}.
\end{eqnarray}
Finally, we can write the dual eigenvalue equation of the Liouvillian as
\begin{eqnarray}
\cL\hat\rho_{k,m}^{(l,n)}=\lambda_{k,m}^{(l,n)}\hat\rho_{k,m}^{(l,n)},\quad
\cL^\dagger\check\rho_{k,m}^{(l,n)}=\lambda_{k,m}^{\ast(l,n)}\check\rho_{k,m}^{(l,n)},
\end{eqnarray}
with the eigenvalues of~\eref{eigenvaluesM} and the right and left eigenvectors given
in~\eref{righteigenL} and~\eref{lefteigenL}, respectively. In our treatment, we have only considered
non-negative values of $l$, however, by including negative values of $l$ according to the following relations
\begin{eqnarray}
\hat\rho_{-k,m}^{(-l,n)}=\hat\rho_{k,m}^{\dagger(l,n)},\\
\check\rho_{-k,m}^{(-l,n)}=\check\rho_{k,m}^{\dagger(l,n)},\
\lambda_{-k,m}^{(-l,n)}=\lambda_{k,m}^{\ast(l,n)}
\end{eqnarray}
one obtains the full eigensystem.

\section{Properties of the damping basis} 
\label{applications}
The formal expressions~\eref{righteigenL} and~\eref{lefteigenL} for the right and left eigenvectors are still written in terms of the superoperators $\cM^{(l,n)}$ and their adjoints. 
This is already an exact solution to the problem, as we have explicitly derived the eigenbasis of $\cM$. 
Therefore, one can use this knowledge
to express the solutions merely in terms of operators. In~\eref{righteigenL} we see that the first superoperator that is applied contains $\cM^{(l,n-1)}$ and we can therefore expand $\hat{\mu}^{(l,n)}_{k,m}$ in terms of its eigenvectors according to
\begin{eqnarray}
\hat{\mu}^{(l,n)}_{k,m}=\sum_{k',m'}c_{k',m'}^{l,k,m}\hat{\mu}^{(l,n-1)}_{k',m'},
\end{eqnarray}
where the coefficients do not depend on $n$ and are given by the scalar products
\begin{eqnarray}
\label{overlap}
c_{k',m'}^{l,k,m}=\Tr\left\{\check{\mu}^{(l,n-1)\dagger}_{k',m'}\hat{\mu}^{(l,n)}_{k,m}\right\}
=C_{k',m'}^{k,m} 
e^{l(\beta^2-\beta^{\ast 2})/2}.
\end{eqnarray}
The remaining trace has already been calculated in the Appendix of Ref.~\cite{Betzholz2014} and reads
\begin{eqnarray}
C_{k',m'}^{k,m}&=\Tr\left\{\check{\mu}_{k',m'}^\dagger D^\dagger(\beta)\hat{\mu}_{k,m}D(\beta)\right\}\nonumber
\\
&=\frac{
	m'!|\beta|^{2(m'-m)+|k'|-|k|}
	(\mbar+1)^{m-m'}e^{i\phi(k'-k)}
}{m!(m'-m-|k|i_-)!(m'-m+|k'|-|k|i_+)!},
\end{eqnarray}
where $\phi=\arg(-\beta)$ and $i_\pm=|k/|k|\pm|k'|/k'|/2$. 
The superoperator can then be applied to its eigenvectors producing the corresponding eigenvalue. Successively iterating this procedure enables us to rewrite the right eigenvectors for $j<n$ as
\begin{eqnarray}
\hat\varrho_{k,m}^{(l,n;j)}=
a_{l,n,j}
\mathop{\sum_{\{k_r,m_r\}}}_{r=j}^{n-1}
\Bigg[
\prod_{s=j}^{n-1}
\frac{\kappa c_{k_s,m_s}^{l,k_{s+1},m_{s+1}}}{\lambda_{k,m}^{(l,n)}-\lambda_{k_s,m_s}^{(l,s)}}\Bigg]\hat{\mu}_{k_{j},m_{j}}^{(l,j)},
\end{eqnarray}
with the definition $k_n,m_n=k,m$. In this notation the sum extends over the set of indices $\{k_r,m_r\}_{r=j}^{n-1}$ specified by the lower and upper line of the sum symbol. Written in this form all superoperators have been eliminated and in the same manner one can derive an expression for the left eigenvectors for $j>n$ according to
\begin{eqnarray}
\check\varrho_{k,m}^{(l,n;j)}=
a_{l,j,n}
\mathop{\sum_{\{k_r,m_r\}}}_{r=n+1}^{j}
\left[\prod_{s=n+1}^{j}\frac{\kappa c_{k_{s-1},m_{s-1}}^{\ast l,k_s,m_s}}{\lambda_{k,m}^{\ast(l,n)}-\lambda_{k_s,m_s}^{\ast(l,s)}}\right]\check{\mu}_{k_j,m_j}^{(l,j)}.
\end{eqnarray}

The right eigenvector associated with the eigenvalue $\lambda_{0,0}^{(0,0)}=0$ represents the single steady state of the optomechanical master equation. It is invariant under the action of $\exp(\cL t)$ and has the form $\hat{\rho}_{0,0}^{(0,0)}=|0\rangle\langle 0|\hat{\mu}_{0,0}$, where $\hat{\mu}_{0,0}=[\mbar/(\mbar+1)]^{b^\dagger b}/(\mbar+1)$ is the thermal-equilibrium density operator. The corresponding left eigenvector is the unity operator and the fact that the real part of all other eigenvalues is negative ensures that every initial condition is damped toward this state as time tends to infinity. 
In Ref.~\cite{Englert2002} it is verified that the right and left eigenvectors of $\cL_{\rm M}$ form a complete biorthogonal basis of operators. Considering the shape of the transform~\eref{asdis}, this directly implies that the right and left eigenvectors of $\cM^{(l,n)}$, for any given value of $l$ and $n$, likewise form such a basis. Since the Fock basis is complete, the right eigenvectors of $\cM$, given in~\eref{eigenvalueeqM}, together with their left counterparts, are a complete basis for operators acting on the full Hilbert space, which also proves the completeness of the eigenbasis of $\cL$. One can furthermore verify the orthonormality by evaluating the scalar product of left and right eigenvectors, given by
\begin{eqnarray}
\label{orthotrace}
\Tr\left\{\check{\rho}_{k,m}^{\dagger(l,n)}\hat{\rho}_{k',m'}^{(l',n')}\right\}=
\delta_{l,l'}\sum_{j=n}^{n'}\Tr\left\{\check{\varrho}_{k,m}^{\dagger(l,n;j)}\hat{\varrho}_{k',m'}^{(l,n';j)}\right\}.
\end{eqnarray}
For $n=n'$ these traces are nothing but scalar products of eigenvectors of $\cM^{(l,n)}$ that reduce to the scalar products of damped oscillator eigenvectors, which are known to be orthonormal~\cite{Briegel1993}. In the case $n\neq n'$ the trace vanishes, as it can be noted
that the scalar products reduce to
\begin{eqnarray}
\Tr\left\{\check{\rho}_{k,m}^{\dagger(l,n)}\hat{\rho}_{k',m'}^{(l',n')}\right\}&=\frac{\delta_{l,l'}}{a_{l,n,n'}}
\mathop{\sum_{\{k_r,m_r\}}}_{r=n+1}^{n'-1}
\Bigg[\prod_{s=n+1}^{n'}\kappa c_{k_{s-1},m_{s-1}}^{l,k_s,m_s}
\Bigg]\nonumber\\
&\times\sum_{j=n}^{n'}
\Bigg[\prod_{s=n+1}^{j}\frac{1}{\lambda_{k,m}^{(l,n)}-\lambda_{k_s,m_s}^{(l,s)}}
\Bigg]\Bigg[
\prod_{s=j}^{n'-1}
\frac{1}{\lambda_{k',m'}^{(l,n')}-\lambda_{k_s,m_s}^{(l,s)}}\Bigg].
\end{eqnarray}
The sum in the second line can be converted into the following form
\begin{eqnarray}
\frac{1}{g}\sum_{j=n}^{n'}
\Bigg[
\prod_{s=j+1}^{n'}(\lambda_n-\lambda_s)
\Bigg]\Bigg[
\prod_{s=n}^{j-1}(\lambda_{n'}-\lambda_s)
\Bigg]=0,
\end{eqnarray}
where we have omitted the other indices and $g$ is the multiplication of all monomials
of the form $\lambda_{n}-\lambda_s$ and $\lambda_{n'}-\lambda_s$.
This can be shown to be zero by expanding in terms of elementary symmetric polynomials~\cite{Macdonald1979}.

\section{Ultra-strong coupling master equation}
\label{strong}
In the ultra-strong coupling regime, $\chi/\nu\gtrsim 1$, dissipative effects are mixed and one has to take into 
account combined decay mechanisms~\cite{Hu2015}. This was noted in Ref.~\cite{Hu2015},
where the dressed-state basis of Hamiltonian~\eref{Hamiltonian} was used in the derivation of the dynamical equation.
The so-called dressed-state master equation (DSME) has the form of~\eref{master}, but with the 
following replacements
\begin{eqnarray}
\cL_{\rm C}=-i\omega \cC[a^\dagger a]+\frac{\kappa}{2}\cD[a]
+\frac{4\chi^2\gamma}{\nu^2\ln\frac{\mbar+1}{\mbar}}\cD[a^\dagger a],
\\
\cL_{\rm M}=-i\nu\cC[b^\dagger b]+\frac{\gamma(\mbar +1)}{2}
\cD[B]+\frac{\gamma \mbar}{2}\cD[B^\dagger],
\end{eqnarray}
where $B=b-\chi a^\dagger a/\nu$. This DSME can also be solved in closed analytical form.
With the aid of an asymmetric displacement of the form~\eref{asdis}, it is possible to bring the system into the form of~\eref{Mtilde}, 
provided the following parameters are chosen
\begin{eqnarray}
\beta=\chi/\nu,\quad
\alpha_{l,n}=-(n+l)\beta,\quad
\beta_{l,n}=-n\beta,\quad
 \eta_l=0.
\label{alfabetas2}
\end{eqnarray}
In this case, one finds that the corrections of the eigenvalues in~\eref{eigenvaluesM} take the following
form
\begin{eqnarray}
\varepsilon_{l,n}=l\beta^2\left[i (2 n +|l|)\nu-\frac{4l\gamma}{\ln\frac{\mbar+1}{\mbar}}\right].
\end{eqnarray}
The imaginary part has the same shape, while the correction to the real part shows a faster growth
with temperature. The strong-coupling damping basis can be completely evaluated with the previous expressions and the mentioned replacements.
It is remarkable that the solution in this case can be obtained in simpler form.

\section{Conclusions}
We presented the analytical diagonalization of the Liouville superoperator generating the
dissipative dynamics of a typical optomechanical system. Exact expressions for the eigenvalues as well as the
eigenvectors were derived. 
Indeed, the
exact solution of the damped harmonic oscillator has been known for decades, 
however, little progress was made in solving more sophisticated problems in this manner or in exploiting
the potential of the original damping basis for specific applications. 
In this work, we have solved the problem of a nonlinear system of coupled harmonic oscillators. Furthermore,
with the case of the master equation in the strong-coupling regime we have presented the first exact solution 
to an eigenvalue problem incorporating combined decay mechanisms for infinite-dimensional systems.
The aim of this work is to extend the understanding of open quantum systems in terms of their analytic solution
and to draw attention back to the usefulness of damping bases.

\ack
R.B. is grateful for financial support of the China Postdoctoral Science Foundation through Grant No. 2017M622398 and for the hospitality of the Instituto de F\'isica of the Benem\'erita Universidad Aut\'onoma de Puebla during the completion of this work. J.M.T. acknowledges support by PRODEP-SEP project 511-6/18-9344. The authors thank Simon B. J\"ager for useful comments. 

\section*{References}
\providecommand{\newblock}{}

\end{document}